\documentclass{optica-article}

\journal{opticajournal} 

\articletype{Research Article}

\usepackage{lineno}
\usepackage[dvipsnames]{xcolor}
\usepackage{upgreek}
\usepackage{graphicx}
\usepackage{tabularx}
\usepackage{hyperref}

\begin{document}

\title{Process-structure-property relationships in subtractive fabrication of silicon nitride microresonators for nonlinear photonics}

\author{Lala Rukh,\authormark{1,2} Gabriel M. Colaci\'{o}n,\authormark{1,2} Franco H. Buck,\authormark{1,2} and Tara E. Drake\authormark{1,2,3,*}}

\address{\authormark{1}Center for High Technology Materials, University of New Mexico, Albuquerque, New Mexico, USA\\
\authormark{2}Optical Science and Engineering Program, University of New Mexico, Albuquerque, New Mexico, USA \\
\authormark{3}Department of Physics and Astronomy, University of New Mexico, Albuquerque, New Mexico, USA}

\email{\authormark{*}drakete@unm.edu}

\begin{abstract*} 
Silicon nitride (SiN) has emerged as a popular platform for nonlinear photonics due to its large bandgap, relatively large nonlinear index of refraction, and high degree of CMOS-compatibility. Nonlinear optical processes in microresonators require both large quality factors to promote build-up of large intracavity powers and sufficient control over resonator dispersion to implement phase-matching, which is generally achieved through careful design and fabrication of the resonator cross-sectional dimensions. 
Additionally, generating dissipative Kerr soliton optical frequency combs requires operation in the anomalous dispersion regime, which calls for such thick layers of SiN as to make achieving accuracy during etching a challenge.
In this work, we investigate the relationship between fabrication process details, including lithography and etching, and optical properties, such as optical loss and resonator dispersion, for two different subtractive fabrication approaches: one utilizing a polymer-based electron beam resist as the protective mask during etching and the other employing a thin metallic etch-mask. Using both approaches, we identify the advantages and disadvantages of polymer and metal etch-masks, determine the origins of optical loss through variations in electron beam lithography parameters, and connect resonator sidewall roughness to optical loss. Finally, we compare frequency comb generation in resonators fabricated using either approach. 

\end{abstract*}

\section{Introduction}

The development of versatile platforms for nonlinear photonics is an important and promising direction in the field of photonic integrated circuits (PICs).
Stoichiometric silicon nitride is a popular material for integrated photonics due to its spectrally wide transparency window \cite{blumenthal_silicon_2018,gaeta_photonic-chip-based_2019}, the absence of two-photon absorption for telecom wavelength \cite{kruckel_optical_2017}, and the potential for planar integration with complementary metal oxide-semiconductors (CMOS) technology \cite{moss_new_2013,levy_cmos-compatible_2010}. Silicon nitride is particularly attractive for nonlinear applications due to a large linear index of refraction ($\mathrm{n \approx 2.0}$ at telecom wavelengths)\cite{luke_broadband_2015} and a nonlinear index of refraction that is ten times higher than fused silica ($\mathrm{n_2=1.9*10^{-19}}$ $\mathrm{m^2/W}$) \cite{ikeda_thermal_2008}, both of which enable efficient nonlinear processes. 

Frequency combs in silicon nitride (hereafter SiN) microresonators are generated via four-wave mixing, a third-order nonlinear effect. The threshold power for four-wave mixing scales as $\mathrm{V/Q^2}$ \cite{PhysRevLett.93.243905}, where V is the mode volume and Q is the internal quality factor of the pumped resonance; thus, resonators with strong optical confinement and low optical loss are necessary for efficient comb generation. Additionally, designing microresonators for optical frequency combs often requires tailoring of the resonator dispersion. For instance, producing a dissipative Kerr soliton (DKS) comb requires anomalous group velocity dispersion in regions where SiN is naturally normally dispersive \cite{scienceaan8083}, and some DKS applications require dispersion control spanning as far as an octave to facilitate stabilization via f-2f self-referencing \cite{drake_terahertz-rate_2019}. Tailoring the dispersion of microring resonators requires control of the waveguide shape and dimensions; for broadband DKS combs pumped at telecom wavelengths, the microring waveguide is typically 600-800 nm tall and twice as wide, with dimensions that are faithfully reproduced at a tolerance of less than 10 nm \cite{briles_interlocking_2018,colacion_design_2024}.
However, silicon nitride is resistant to both chemical and physical etching, and etching requires a protective etch-mask that is either more etch-resistant or much thicker than the nitride layer so that it outlasts the nitride removal, leading to inevitable challenges in achieving the required dimensions.

Achieving dimensional control without sacrificing quality factors can prove difficult. Etch-masks are most often developed lithographic resist, which offers the advantage of fewer total steps in the fabrication process \cite{ye_high-q_2019,li_vertical_2013,xuan_high-q_2016}. However, resist layers that are thick enough to survive etching come with various potential drawbacks, including aspect ratio dependent etching, etch-depth nonuniformity, and sloped sidewalls, all of which contribute to poor dimensional control. An alternative approach is to use masks that are highly etch-resistant, such as dielectric (e.g., silica) or metallic masks \cite{lim_development_2011,colacion_low-loss_2025,ji_methods_2021}. Because of their high etch-resistivity, these masks can be thin enough to reduce RIE lag and produce waveguides with uniform etch depth and vertical sidewalls. While these masks can offer increased dimensional control, they also require additional processing steps, are typically not as smooth as polymer-based resist, and/or can be difficult to remove post-etch.

This work is a dedicated study of the connections between fabrication process details (specifically, lithography and etching) and optical properties (quality factors/optical loss and resonator dispersion/comb spectral design) for subtractively fabricated SiN microring resonators. We investigate two general approaches to subtractive fabrication of SiN waveguides. The first fabrication process uses a thick layer of polymer-based electron-beam resist as the protective mask during the etching step \cite{li_vertical_2013} and represents a typical SiN subtractive fabrication flow. The contrasting approach uses a thin, metallic etch-mask \cite{lim_development_2011,colacion_low-loss_2025} that exploits the durability of metal during reactive ion etching (RIE). This process uses metal lift-off to transfer the waveguide pattern from electron-beam resist to the metal mask. (We have published the details of our metal mask fabrication process in reference \citenum{colacion_low-loss_2025}.) 
In either process, we fabricate microrings with tight-bending radii in SiN from the same deposition run; in this way, our comparison emphasizes the optical mode interaction with the waveguide sidewalls and keeps any material absorption common to all devices. 
We identify a tradeoff between lower optical loss achieved using the polymer mask and more vertical sidewalls and uniform etching depths realized using the metal mask. We probe the origin of optical loss in either process through both an experimental investigation of electron beam lithography parameters (e.g. beam diameter and grid spacing) and simulations connecting resonator sidewall roughness to loss via the Payne-Lacey model 
\cite{Payne_1994}. Finally, we show that both fabrication processes yield microresonators capable of supporting soliton frequency combs and octave (or near-octave) comb bandwidths.

\section{Fabrication Details for Polymer and Metal Etch-Mask Process Flows}

For this series of experiments, we fabricate identical sets of silicon nitride microring resonators evanescently coupled to straight sections of the access waveguides (i.e., point couplers). Each microring is designed with cross-sectional dimensions to produce a group velocity dispersion that is anomalous in the telecom C-band and has higher order terms that produce dual, harmonic dispersive waves at roughly 1000 and 2000 nm when the resonator is pumped near 1550 nm \cite{Brasch_2016,briles_interlocking_2018}. The critical (i.e., smallest) dimension of these devices is the gap between the waveguide and microring, which we vary from 300 nm to 800 nm in order to be able to distinguish propagation loss in the waveguide from coupling loss. We purposefully fabricate rings with small radii (R = 23.3 $\upmu$m) for two important reasons. First, a large free spectral range (FSR) is necessary for achieving octave-spanning spectra, which is critical for our investigation of higher order dispersion terms and the presence of dispersive waves. Second, the tight bending radius will enhance optical mode interaction with the waveguide sidewall, which will increase the losses associated with sidewall physical properties \cite{ji_methods_2021}. This allows us to focus our investigation on the loss contributions most directly associated with subtractive fabrication. Additionally, we expect optical loss due to absorption to be similar for all resonators in this study, as we keep all process steps not dependent on etch-mask type identical for the two methods.

All microresonator devices in this work are fabricated using one of two general fabrication processes, one of which uses a polymer-based lithographic resist as the protective mask layer during etching, and one which uses a thin layer of chromium metal as the protective etch-mask. For both polymer- and metal-mask process flows, fabrication starts with 3-inch bare Si wafer on which we thermally grow 3 microns of silica. We deposit 600 nm of stoichiometric SiN via low pressure chemical vapor deposition (LPCVD) in a careful multi-step process that protects the SiN from cracking and implements high temperature annealing to eliminate N-H bonds \cite{Henry1987}. The lithographic patterning and inductively coupled plasma reactive ion etching (ICP-RIE) steps differ significantly between the polymer mask and metal mask processes, and the parameters of the e-beam lithography is further varied within each process type. (Overall, we report on seven unique sets of process parameters, three using a polymer etch-mask and four using a metal etch-mask.) After etching and mask removal, we create facet-released edge-couplers at either end of our access waveguides via a combination of photolithography and etching \cite{ji_methods_2021}. Further details about SiN deposition, microring fabrication, and the facet release process are explained in Sections 1-3 and Fig. S1 of the Supplemental Document. 

\begin{figure}[t]
    \centering
    \includegraphics[width=1\linewidth]{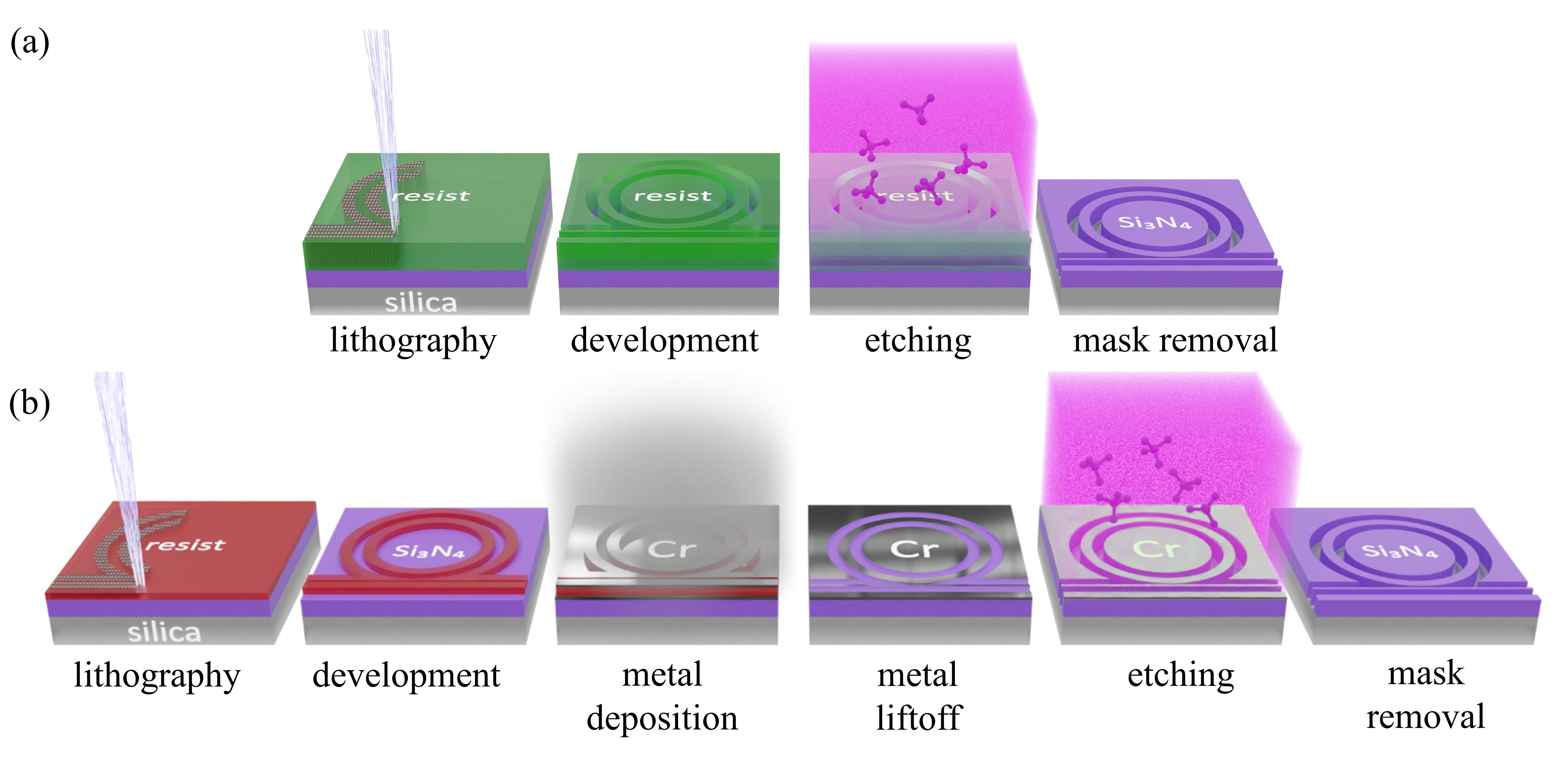}
    \caption{Schematics for polymer and metal etch-mask processes. (a) The polymer mask process starts by coating a thick layer of positive-tone resist, which is patterned via electron beam lithography. The exposed resist is developed and is used as the protective etch-mask for inductively coupled plasma reactive ion etching (ICP-RIE). The remaining resist is removed after etching. (b) The metal mask process starts by coating the nitride with a thin layer of negative-tone resist. The resist is exposed using electron beam lithography using the same pattern as in panel a. The exposed samples are then developed, and a thin layer of chromium metal is deposited via electron beam evaporation. The samples are soaked overnight in a solvent to remove all resist (and to ``lift-off'' the metal that is on top of resist). This transfers the negative of the resist pattern to the metal, which then serves as the protective etch-mask during ICP-RIE. The remaining metal is removed via chromium etchant, leaving behind the SiN resonators. This panel is reproduced from reference \citenum{colacion_low-loss_2025}. In both panels, the $\mathrm{CF_{4}}$ crystal structure is adapted from reference \citenum{PubChem}. } 
    \label{fig:fig1}
\end{figure}

\begin{figure}[ht!]
    \centering
    \includegraphics[width=1\linewidth]{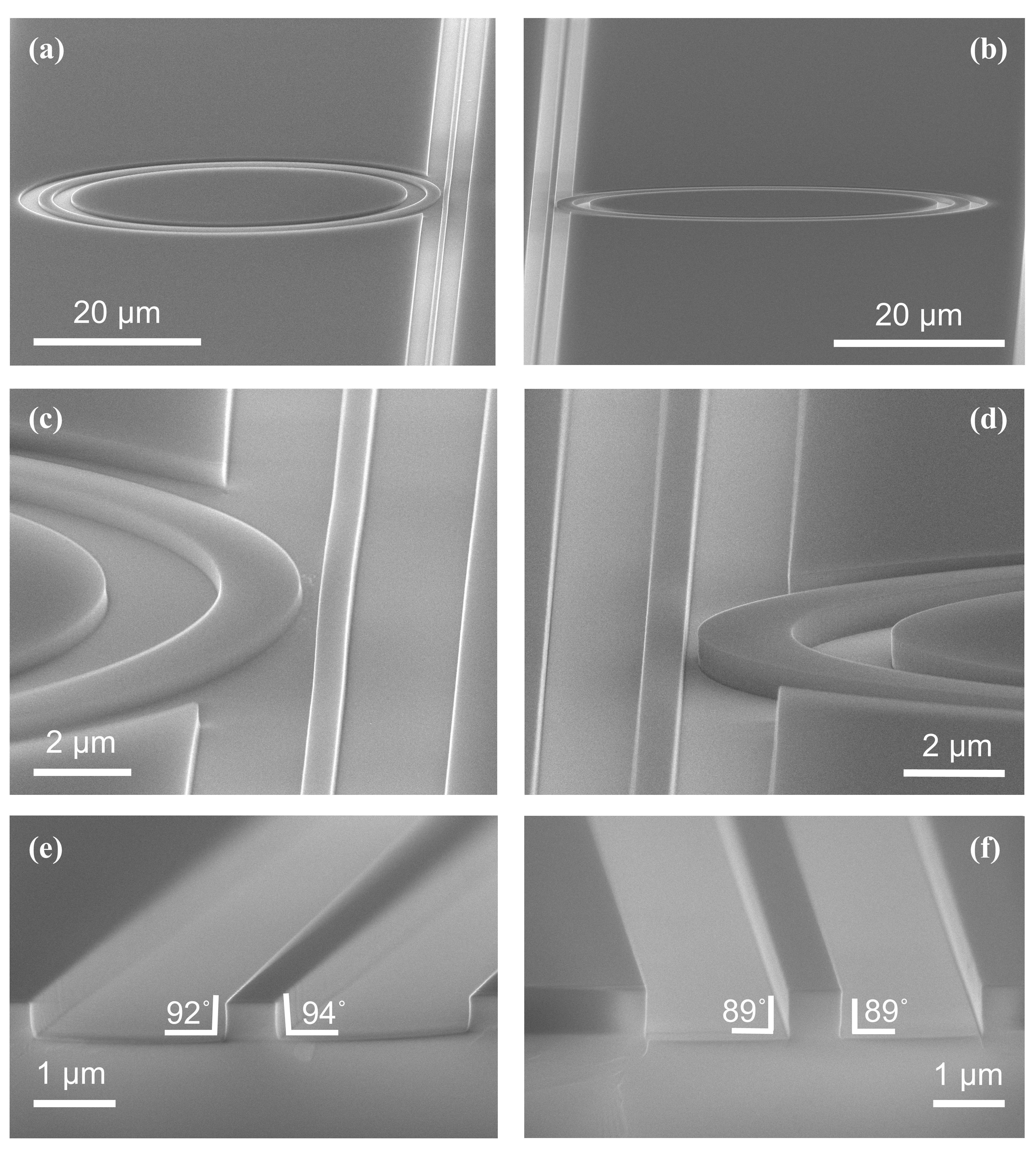}
    \caption{SEM micrographs of SiN resonators fabricated using polymer and metal mask processes. We show microring resonators with access waveguides fabricated using a (a) polymer and (b) metal protective etch-mask. (c) Enlarged image of waveguide-resonator gap (critical dimension) for a polymer mask device. (d) Enlarged image of waveguide-resonator gap fabricated using metal etch-mask. (e) Cleaved access waveguide fabricated using a polymer etch-mask. The waveguide sidewalls display slight sloping and bowing. (f) Cleaved access waveguide fabricated using metal etch-mask with a near-vertical sidewall profile.} 
    \label{fig:fig2}
\end{figure}

Figure \ref{fig:fig1} portrays the fabrication steps associated with e-beam lithography and ICP-RIE for the polymer and metal mask process flows. The polymer mask process starts with spin coating SiN with a 1 $\mathrm{\upmu m}$ thick layer of ZEP520A, a positive-tone electron beam resist. We use ZEP520A because of its high resolution and dry-etch resistance \cite{ZEPtechsheet}. 
The exposed resist is developed using n-Amyl Acetate. This resist serves as the etch template for the silicon nitride waveguides. The metal mask process begins with coating a 250 nm thick layer of ma-N 2403, a negative-tone electron beam resist. We use ma-N 2403 due to its increased thermal resistance and high resolution \cite{maNtechsheet}.
The exposed sample is then developed using MF 319. After development, we deposit a 50 nm layer of chromium using electron beam physical vapor deposition (EBPVD). During metal lift-off, the samples are soaked in \textit{N}-Methyl-2-pyrrolidone (NMP) to remove the resist and transfer the pattern in metal, which will serve as the etch-mask template \cite{colacion_low-loss_2025}. Both resists are exposed using a JEOL JBX 6300-FS electron beam lithography system with an acceleration voltage of 100 kV. In our experiments, we vary both the spot size of the electron beam at the resist surface (controlled via beam current) and the spacing of the discrete grid (shot pitch) which forms the pattern we lithographically print.
This allows us to investigate the connection between optical loss and lithography parameters.

We etch all devices using a Plasma-Therm SLR system and a mixture of carbon tetrafluoride ($\mathrm{CF_{4}}$), argon ($\mathrm{Ar}$), and oxygen ($\mathrm{O_{2}}$). We optimize the etch recipes for either process flow independently to enhance etch-depth uniformity and minimize RIE lag. We keep the chemistry of the etch the same for both processes, but the physical etch parameters including the bias RF power and DC bias are set to balance the trade-off between etch selectivity and directional etching. We define etch selectivity as the ratio of the etch rate of the protective mask to the etch rate of the target material. 
The etch selectivity for the developed resist compared to the SiN is 1.4:1, while for metal mask, the etch selectivity is 1:30. Due to the poor etch selectivity of the polymer mask process, we choose a low bias RF power compared to the etch recipe for the metal mask. 

We use scanning electron microscope (SEM) images to determine the best etching parameters for either mask type. Figure \ref{fig:fig2}(a) and \ref{fig:fig2}(c) shows SEM images of fabricated resonators using a polymer mask, e-beam current of 1 nA, and step size of 8 nm. The use of a mask thicker than the SiN in the polymer mask recipe creates deep and narrow trenches that cause ion depletion and inhibit the etch process and makes this process more prone to etch-depth nonuniformity and aspect ratio dependent etching \cite{Gottscho1992}. 
Figure \ref{fig:fig2}(b) and \ref{fig:fig2}(d) show SEM images of fabricated resonators using a metal mask, e-beam current of 200 pA, and 2 nm shot pitch. The zoomed in image of waveguide-resonator gap in Fig. \ref{fig:fig2}(d) shows a clean and uniform etch depth and no evidence of RIE lag. The dispersion of a resonator is sensitive to sidewall angle; small variations in the angle can dramatically change the higher-order dispersion terms, affecting the bandwidth of the generated light. Figures \ref{fig:fig2}(e) and \ref{fig:fig2}(f) show the difference in waveguide sidewall angle between the two processes, with the polymer-mask processed waveguide exhibiting a slight but clear slant of roughly 93° and slight concavity, while the metal-mask processed waveguide produces vertical sidewalls (89°) without curvature. We estimate our ability to measure sidewall angle in these SEM images at 1° precision, with charging effects making more precise measurements difficult. We have observed that increasing the bias RF power in the etcher is correlated with waveguide sidewall orthogonality, and also that higher bias power leads to worse etch selectivity for the polymer mask. We attribute the difference in sidewall angle partially to the lower bias power in the polymer mask etch recipe chosen to balance these two effects.

\section{Characterization of Intrinsic Optical Loss: Quality factor measurements}
\label{sec:Qfactor}

The requirement of low threshold for four-wave mixing necessitates photonic devices with low optical loss. We investigate the optical loss of fabricated resonators by measuring the total, or loaded, quality factors of individual resonances for multiple microresonators. 

Quality factor is defined as the ratio of energy stored in a cavity to the energy lost per cycle. It is equivalent to the ratio of the resonance frequency to the frequency width of the resonant mode \cite{Heebner2008,Tobing2010,Haus1984}. In the weak coupling limit, where both propagation loss and coupling loss is small, the loaded quality factor is separable into the these two categories of loss:  
\begin{equation}
\mathrm{1/Q_{loaded}=1/Q_{int}+1/Q_{c}} .
\label{eq:Q}
\end{equation}
The intrinsic or internal quality factor, $\mathrm{Q_{int}}$, originates from internal losses within the resonator such as scattering and absorption, while the coupling or external quality factor,  $\mathrm{Q_{c}}$, represents the out-coupling of light from the microring resonator to the access waveguide. 

To measure the quality factor, we fabricate a series of resonators with waveguide-resonator coupling gaps from 300 nm to 800 nm. Each resonator should have similar instrinsic losses, while coupling is reduced and $\mathrm{Q_{c}}$ increased for larger gaps. For each resonator in the series, we measure the 16 resonant TE modes that span telecom wavelengths from 1510 nm to 1630 nm. We record the optical frequency of the mode, the on-resonance transmission of the laser, T, and the frequency width of the mode at the halfway point between maximum and minimum transmission (FWHM).  
We extract $\mathrm{Q_{int}}$ from the total quality factor using 
\begin{equation}
\mathrm{Q_{int}=2*Q_{loaded}/(1\mp\sqrt{T})}
\label{eq:coupling}
\end{equation}
\noindent
for overcoupled (-) and undercoupled (+) resonators, respectively \cite{Borselli2005}.

\begin{figure}[htbp]
\centering
\includegraphics[width=0.9 \linewidth]{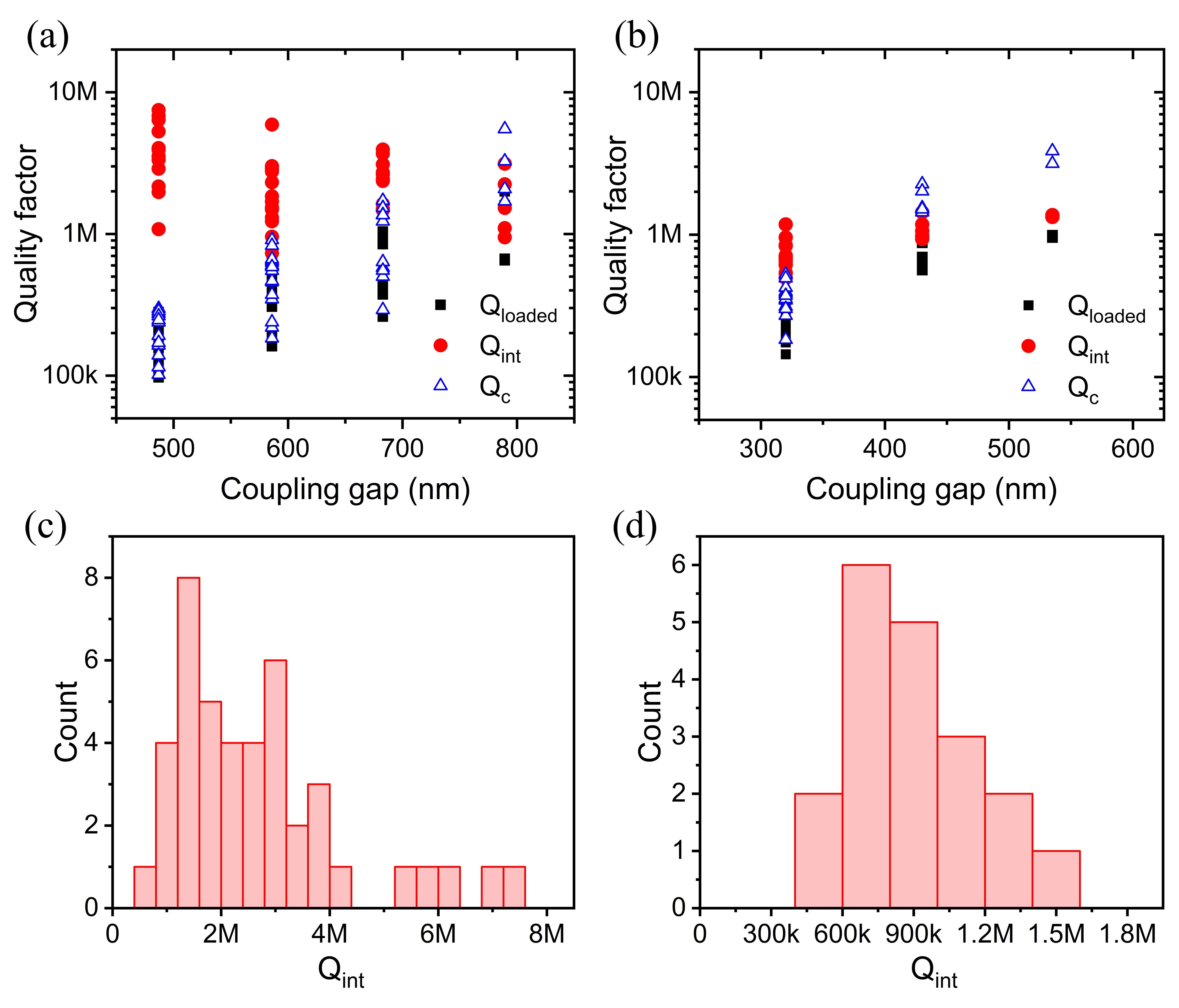}
\caption{Measured quality factors of polymer and metal mask devices. Loaded, intrinsic, and coupling quality factors as a function of coupling gap for (a) polymer and (b) metal mask process. Histograms of intrinsic quality factors measured in (c) polymer and (d) metal mask devices. (Panels b and d are adapted from reference \citenum{colacion_low-loss_2025}.) }
\label{fig:fig3}
\end{figure}

Figure \ref{fig:fig3}(a) shows intrinsic, coupling, and loaded quality factors for a series of devices fabricated using a polymer etch-mask with consistent process details (beam current of 200 pA and shot pitch of 4 nm). As expected, the coupling (and the loaded) quality factors increase as the coupling gap increases, while the intrinsic quality factor remains relatively flat. For this series of resonators, we report 43 resonances and an average $\mathrm{Q_{int}}$ of 2.7(2) million and a standard deviation of 1.6M. The variance in quality reflects slight statistical differences in loss for different longitudinal modes within the TE$\mathrm{_{00}}$ family. We sometimes observe a larger spread in our calculation of $\mathrm{Q_{int}}$ variance for the most overcoupled devices; this is an expected consequence of the increased sensitivity of the overcoupled version of equation \ref{eq:coupling} to small errors in transmission as T approaches 1. We display the distribution of resonance intrinsic quality factors for these resonators in Fig. \ref{fig:fig3}(c). 
Similarly, Fig. \ref{fig:fig3}(b) and (d) show quality factors recorded for a series of devices fabricated using metal etch-masks (beam current 200 pA and shot pitch 2 nm). 
The average intrinsic Q for the 19 resonances reported in this series is 0.90(7) million with a standard deviation of 290k.

There are a few general trends we observe in Fig. \ref{fig:fig3} that are consistent for all data presented in this paper. First, we observe that resonators fabricated via the polymer mask process typically have quality factors that are significantly larger than resonators fabricated using metallic etch-masks. Also, the coupling gap width where we find critical coupling (i.e., where $\mathrm{Q_{int}} = \mathrm{Q_{c}}$ and $\mathrm{T = 0}$) is larger for the polymer mask devices (between 700 and 800 nm for this data) than for the metal mask devices (between 300 and 400 nm); this comes in part from the larger $\mathrm{Q_{int}}$ values for the former. (Reported gap sizes are determined via post-fabrication SEM measurements of each device in question.) We also apply selection rules in our reporting of $\mathrm{Q_{int}}$. In this data, we only include resonances with transmissions that are less than 90\%, the level at which we feel comfortable that small defects in the access waveguide causing etaloning and elliptical polarization will not create excessive error in T. We also exclude any resonances with perceivable scattering doublet splittings. However, although we do not include them in any quality factor measurements, we do investigate doublets in Section \ref{sec:doublets}, as their properties can provide information about sidewall scattering. Finally, we do not include any resonant modes that are hybridized with modes from another mode family (``mode crossings'', typically with the TM$\mathrm{_{00}}$ and higher order TE families).
 
We extract propagation loss, $\mathrm{\alpha}$, from intrinsic quality factor using the relation
\begin{equation}
\alpha=\frac{\lambda_o}{\mathrm{Q_{int}}*\mathrm{R}*\mathrm{FSR}},
\end{equation}
where $\mathrm{R}$ is the radius of the microring, $\mathrm{FSR}$ is the resonator free spectral range (in meters), and $\mathrm{\lambda_o}$ is the wavelength of the resonance \cite{rabiei_polymer_2002}. 
The average propagation loss for the above polymer mask microrings is 0.14 dB/cm, while for the metal mask microrings, the average propagation loss is 0.42 dB/cm. 
In the next sections, we present experiments to determine the origin of intrinsic optical loss within our microrings and to better understand the difference in propagation loss between the two process flows. In particular, we focus on building connections between fabrication parameters (lithography parameters, etch-mask properties), physical microring features (e.g., sidewall roughness, waveguide rectangularity), and optical properties (loss, backscattering) through experiments in which we record quality factors and scattering doublet properties as a function of process parameters as well as through modeling of scattering loss given sidewall roughness.

\section{Investigating the Origins of Optical Loss}

\subsection{Lithography parameter study}

\begin{table}[!ht]
\caption{Average $\mathrm{Q_{int}}$ and propagation loss as a function of etch-mask type and electron beam lithography parameters.}
\centering
\begin{tabularx}{0.95\textwidth}{ >{\centering\arraybackslash}X  >{\centering\arraybackslash}X  >{\centering\arraybackslash}X  >{\centering\arraybackslash}X  >{\centering\arraybackslash}X  > 
{\centering\arraybackslash}X  >{\centering\arraybackslash}X  >{\centering\arraybackslash}X}
\hline

\ Sample & Mask & Beam current & Shot pitch & Avg. $\mathrm{Q_{int}}$ & SE of mean & Std. dev. & Propa-gation loss \\
 &  & (nA) & (nm) &  &  &  & (dB/cm) \\
\hline
S1 & Polymer & 0.2 & 4 & 2.7M & 200k & 1.6M & 0.14\\
S2 & Polymer & 0.2 & 8 & 2.8M &  200k & 1.3M & 0.13\\
S3 & Polymer & 1 & 8 & 1.6M & 100k & 900k & 0.23\\
H1 & Metal & 0.2 & 2 & 900k & 70k & 290k & 0.42\\
H2 & Metal & 0.2 & 6 & 1.10M & 70k & 250k & 0.34\\
H3 & Metal & 1 & 6 & 1.0M & 100k & 500k & 0.38\\
H4 & Metal & 1 & 12 & 990k & 50k & 230k & 0.38\\
\hline
\end{tabularx}
\label{tab:loss}
\end{table}

Although many fabrication processes can be responsible for waveguide optical loss, in this work we focus on investigating the influence of the lithography and etching steps. We build our understanding of these connections by varying the diameter and grid spacing of the individual electron beam ``shots'' used to write our pattern during lithography for either of our two etching processes. We control e-beam diameter and spacing by altering the beam current and machine shot pitch, respectively. Table \ref{tab:loss} presents $\mathrm{Q_{int}}$ and propagation loss for microresonators fabricated from identical patterns and using a range of beam size and spacings for both etch-mask types. (For these data, we use the same selection rules as detailed in Section \ref{sec:Qfactor}.)

The current of the electron beam sets the spot size of the beam before it enters the resist layer. We use beam currents of either 200 pA or 1 nA, which have spot sizes at their focii of 5.8 nm and 7.5 nm, respectively. Interestingly, we observe that decreasing the beam current makes a big difference for the polymer mask process (increasing the quality factor from 1.6(1)M to 2.8(2)M), while making no difference for the metal mask process (1.0(1)M using the larger current/larger beam diameter, compared to 1.10(7)M). It is possible that increasing the total number of electrons poses more of a potential issue when this beam must travel through the 1.0 micron-thick ZEP resist than in the 250 nm-thick ma-N resist, giving rise to a dependence in the first case but not the second. 

We also investigate the change in $\mathrm{Q_{int}}$ for varying shot pitches. For these experiments, one might consider the shot pitch relative to the electron beam diameter. For example, samples S1 and H1 represent shot pitches that are significantly smaller than the beam size, and thus individual shots will have significant overlap. On the other hand, samples S2 and H4 use a shot pitch that is larger than the beam size, and so we expect individual shots to have no overlap and potentially to have unexposed resist between them. Surprisingly, we saw no significant change in quality factors for either mask type when the shot pitch was varied and other parameters kept the same. Given the beam current results, it is likely that Coulomb scattering within the thick ZEP resist layer plays some role here, potentially making the shot pitch change superfluous in this process \cite{Zhou2006}.

The difference in total quality factor between the two process, however, is not fully explained by these results. From this we assume that the source of excess loss in the metal etch-mask devices stems from a process step not investigated here; potentially, this is the metal deposition itself \cite{colacion_low-loss_2025}. If the metal deposition is leading to significantly increased roughness in the waveguide sidewalls, that would also explain the lack of change seen in the metal mask device quality factors for widely varied lithography parameters. (The data presented in this section are shown in more detail in Section 4 and Fig. S2 and S3 of the Supplemental Document.)

\subsection{Analysis of scattering doublets}
\label{sec:doublets}

The experiments presented here are designed to investigate optical loss originating from scattering rather than absorption. All devices are made from the same SiN deposition run and have undergone identical annealing procedures and, so, presumably have identical absorptive losses. Additionally, we fabricate THz-FSR resonators, which have tight bending radii such as have been shown to limit quality factors compared to similar resonators with smaller FSRs, presumably due to increased interactions of the optical mode with the sidewalls \cite{ji_methods_2021}. In this section, we investigate doublet splitting of the optical resonances, which provides direct evidence of loss due to scattering.

An ideal ring resonator system is unidirectional, but any nonideality can excite the mode in opposite direction via backscattering. The coupling between forward and backward propagating modes lifts the degeneracy and causes a splitting of the resonance frequency \cite{li_backscattering_2016,morichetti_roughness_2010}. Backscattering can occur due to many physical contributions, including electric field interaction with the sidewalls and sidewall roughness. The magnitude of resonance splitting depends on the magnitude of backscattering and of the coupling between forward and backward propagating modes. Fabrication-induced scattering is dependent on the length of the waveguide and the sidewall quality. The waveguide length is the same for all devices that are studied for doublet analysis; therefore, resonance splitting is directly related to fabrication-induced sidewall roughness \cite{morichetti_roughness_2010}. 

\begin{table}[!htb]
\caption{Average doublet splitting and doublet prevalence as a function of mask type and electron beam lithography parameters.}
\centering
\begin{tabularx}{0.95\textwidth}{ >{\centering\arraybackslash}X  >{\centering\arraybackslash}X  >{\centering\arraybackslash}X  >{\centering\arraybackslash}X  >{\centering\arraybackslash}X  > 
{\centering\arraybackslash}X  >{\centering\arraybackslash}X  >{\centering\arraybackslash}X}
\hline
\ Sample  &  Mask  & Beam current  & Shot pitch & Avg. doublet splitting  &  SE of mean &  Std. dev.& Doublet prevalence  \\
 &  & (nA) & (nm) & (MHz) & (MHz) & (MHz) & (\%) \\
\hline
S1 & Polymer & 0.2 & 4 & 210 & 20 & 80 & 27\\
S2 & Polymer & 0.2 & 8 & 180 & 10 & 50 & 27\\
S3 & Polymer & 1 & 8 & 180 & 20 & 70 & 21\\
H1 & Metal & 0.2 & 2 & 280 & 20 & 110 & 62\\
H2 & Metal & 0.2 & 6 & 300 & 20 & 140 & 75\\
H3 & Metal & 1 & 6 & 230 & 30 & 90 & 62\\
H4 & Metal & 1 & 12 & 320 & 30 & 150 & 65\\
\hline
\end{tabularx}
\label{tab:doublets}
\end{table}

Table \ref{tab:doublets} summarizes the prevalence and resonance splitting of doublets observed in devices with different lithography conditions and etch-mask material.
We first compare the average splitting and prevalence of polymer and metal mask devices. For all series of resonators investigated, we observe both larger doublet splitting and a greater percentage of modes that are split for the metal mask devices than for those fabricated with polymer masks. This reflects our observation of increased loss for these devices, and further confirms our assumption that this loss is due to scattering originating from sidewall roughness. We speculate that the granularity of the chromium mask layer or the edge roughness of the metal mask after lift-off may be transferring to striations in the resonator sidewalls during etching. In general, we may expect the edge of the polymer mask to be smoother due to the nature of polymer materials. Within each mask type, we do not see that average doublet splitting and percentage of doublet modes has significant dependence on lithography parameters.

\subsection{Modeling sidewall roughness}

\begin{figure}[htpb]
\centering
\includegraphics[width=0.8 \linewidth]{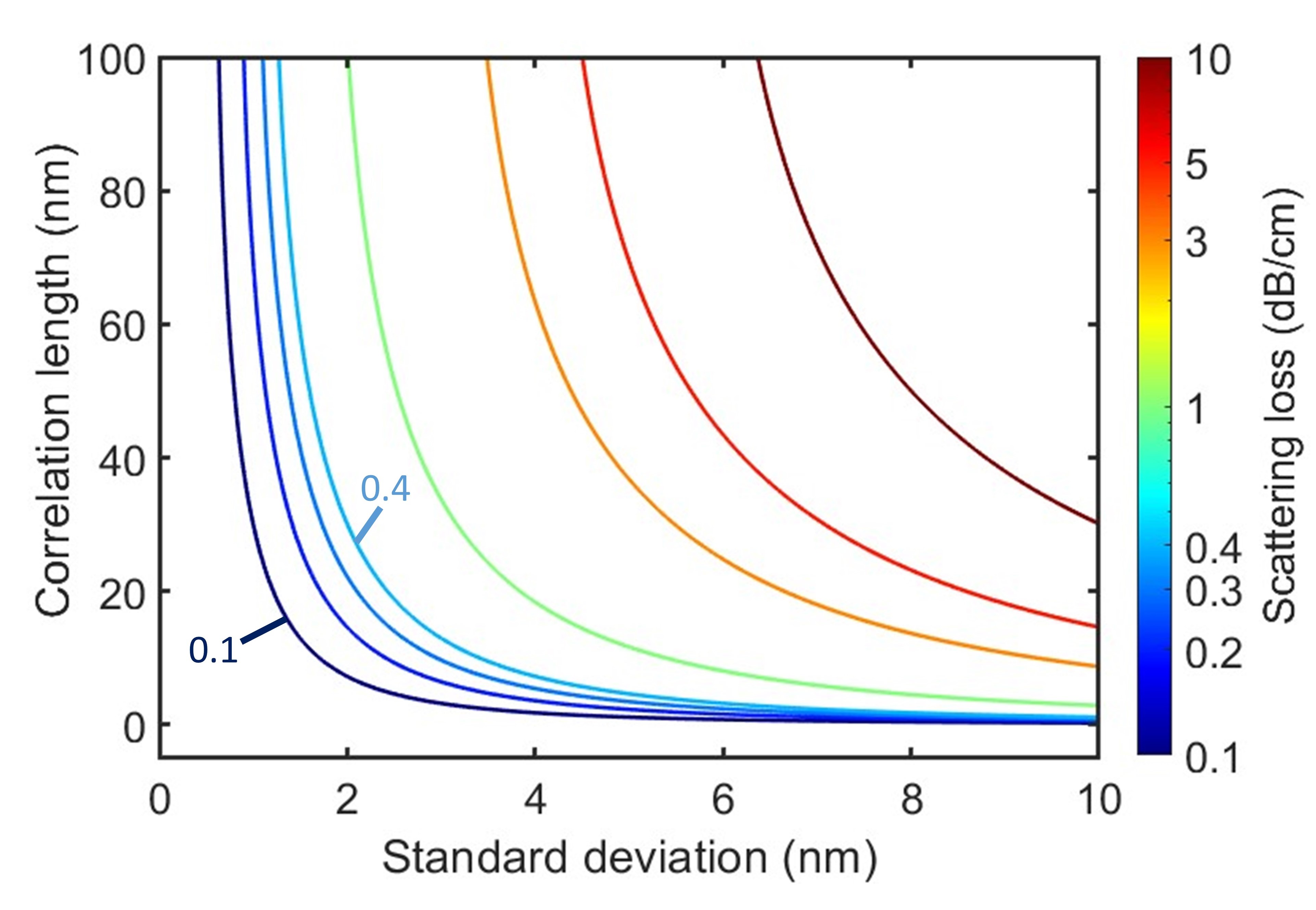}
\caption{Contour map of scattering-induced optical propagation loss as a function of standard deviation ($\mathrm{\sigma}$) and correlation length ($\mathrm{L_{C}}$) of sidewall roughness using a 2-dimensional simulation of a straight waveguide of width 1700 nm at wavelength 1550 nm.}
    \label{fig:contour_plot}
\end{figure}

The presence of strong backscattering evidenced by the prevalence of resonance splitting supports our hypothesis that sidewall roughness is a major contributor to the total propagation loss in our resonators. To connect the loss we observe to the physical properties of the sidewalls, we implement an analytic form for waveguide sidewall roughness-induced scattering loss described by the Payne-Lacey model \cite{Payne_1994,Grillot_2004}. In this model, the sidewall roughness is described by an exponential autocorrelation function which yields the analytic form for the scattering loss coefficient,
\begin{equation}
\mathrm{\alpha_{scatter}=4.34\frac{\sigma^2}{\sqrt{2}\,k_{0}\,n_{c}\,d^4}*g*f}
\label{eq:Payne-Lacey}
\end{equation}
where $\mathrm{k_{0}}$ is the wavevector at our wavelength of interest (1550 nm), $\mathrm{n_{c}}$ is the refractive index of the SiN waveguide, d is the half-width of the waveguide, and $\mathrm{\sigma}$ is the standard deviation of the waveguide half-width (i.e., sidewall roughness) perpendicular to the propagation direction. Both functions g and f depend on the waveguide geometry and materials, and f includes an additional parameter, $\mathrm{L_{C}}$, the correlation length of the roughness in the direction of light propagation. (Section 5 in the Supplemental Document provides more details on the Payne-Lacey model.) 
Expected propagation loss as a function of sidewall roughness parameters, $\mathrm{\sigma}$ and $\mathrm{L_{C}}$, for a SiN waveguide of width 1700 nm and air cladding are shown in Fig. \ref{fig:contour_plot}.

Comparing the propagation loss of samples S1 (0.14 dB/cm) and H1 (0.42 dB/cm) via the associated contour lines, we see a distinct difference in the predicted degree of roughness. For example, for a chosen value of $\mathrm{L_{C}}$ in the range between 20 and 90 nm, which is a reasonable range for electron beam lithographically patterned waveguides \cite{Barwicz2003,Bauters2011,roberts2022measurements}, the sidewall standard deviation typically doubles when scattering loss increases from 0.1 to 0.4 dB/cm. Even if $\mathrm{L_{C}}$ is smaller for the metallic mask devices than the polymer mask devices, in this range, we still expect an increase in $\mathrm{\sigma}$ between the two processes. We also observe that, for reasonable $\mathrm{L_{C}}$ values, $\mathrm{\sigma}$ is likely not larger than 3 nm for either process. This is consistent with our observations that sidewall roughness is too small to be measured via SEM images, which typically cannot measure corrugations below 10s of nm \cite{roberts2022measurements}.

\section {Frequency Comb Generation}

\begin{figure}[h]
\centering
\includegraphics[width=0.75 \linewidth]{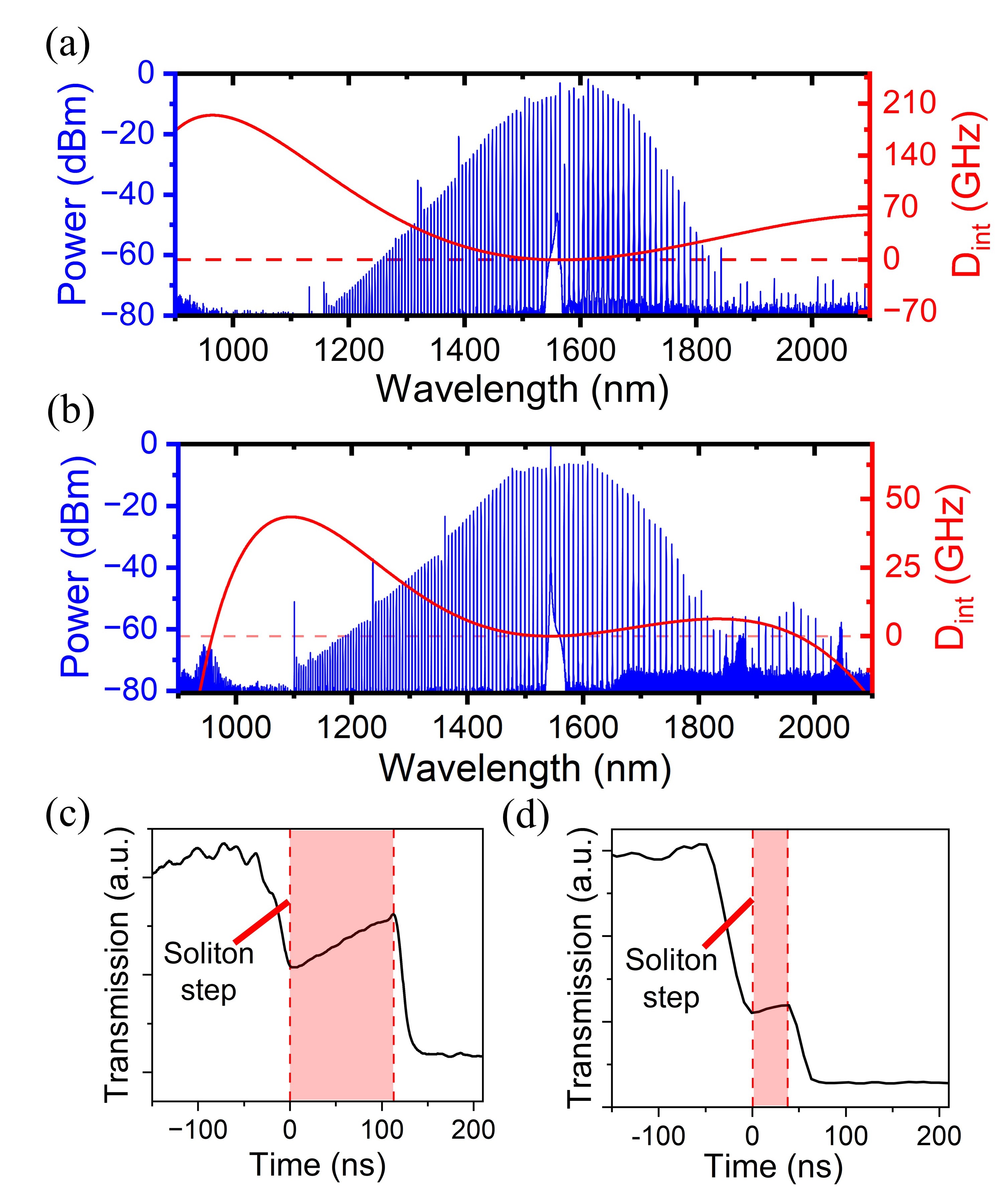}
\caption{Generation of modulation instability (MI) combs and soliton steps using SiN ring resonators fabricated using the polymer and metal mask processes. (a) MI comb generated using a device from sample group S3 with on-chip power of 537 mW and $\mathrm{F^2}$ = 28. The integrated dispersion (red line) is found through FEM simulations of a resonator with the measured dimensions and sidewall angles of the pumped device. (b) MI comb generated using a device from sample group H1 with on-chip power of 195 mW and $\mathrm{F^2}$ = 30. The simulated integrated dispersion (red line) has zero crossings that correspond to the two observed dispersive waves on either end of the spectrum. (c,d) Short-lived solitons (soliton steps) obtained by sweeping the pump laser across a resonance in rings from groups S2 ($\mathrm{Q_{int} = 2.3M}$, 51 mW pump laser at 1541 nm, $\mathrm{F^2 = 31}$) and H1 (same resonator panel b, pumped at $\mathrm{F^2 = 12}$). (Panels b and d are adapted from reference \citenum{colacion_low-loss_2025}.)}
\label{fig:MIcomb}
\end{figure}

To investigate the dispersions of our fabricated devices, we generate broadband combs via cascaded four-wave mixing at large pump laser powers (MI combs). 
Figures \ref{fig:MIcomb}(a) and \ref{fig:MIcomb}(b) show the combs generated in two microresonators fabricated using the polymer and metal mask processes, respectively.
In Fig. \ref{fig:MIcomb}(a), using a device from sample group S3, we send 537 mW of laser power through the access waveguide to pump a resonance at 1564 nm with $\mathrm{Q_{int} = 960k}$. The resonator cross-sectional dimensions were measured post-etching (via SEM images) to be 1689 nm (width) by 617 nm (height). The threshold power for four-wave mixing in the pumped resonance is 19 mW.
In Fig. \ref{fig:MIcomb}(b), we pump a device from the H1 series with 195 mW at 1543 nm. This microring was measured to have a cross section of 1730 nm by 582 nm, and the pumped mode has a $\mathrm{Q_{int}}$ of 1.2M and threshold power of 6.5 mW. The ratio of pump power to threshold power, $\mathrm{F^2=P_{wg}/P_{th}}$, is kept similar for both combs, roughly 28 for Fig. \ref{fig:MIcomb}(a) and 30 for Fig. \ref{fig:MIcomb}(b).

We observe broadband light generation in both devices. We find that the comb generated in the metal-mask process device is more broadband than the comb generated in the polymer-mask process device. Furthermore, the comb in the metal mask device has dual, harmonic dispersive waves on either end of the spectrum, as designed \cite{Brasch_2016}. Dispersive waves  (DWs) occur when the modal walk-off from equal spacing due to dispersion matches the modal walk-off at the pumped wavelength. In the integrated dispersion, $\mathrm{D_{int}(\omega) = \omega_{\mu} - \omega_0 - \mu*FSR}$, where $\mathrm{\omega_{\mu}}$ is the frequency of the comb mode of index $\mu$, and $\mathrm{\omega_{0}}$ is the frequency of the pumped mode, DWs are positioned at the $\mathrm{D_{int} = 0}$ crossings \cite{scienceaan8083}. Considering $\mathrm{D_{int}}$ as the Taylor expansion of the modal dispersion about $\mathrm{\mu = 0}$, one can see that DWs are related to the higher order terms in the dispersion (i.e., beyond quadratic). As such, the presence and position of DWs are very sensitive to small changes in waveguide geometry, making them a good probe of small dimensional deviations from design \cite{briles_interlocking_2018}.

We use finite element method (FEM) simulations of the two resonators to find the expected dispersion in each. $\mathrm{D_{int}}$ in Fig. \ref{fig:MIcomb}(b) (red line) shows zero crossings at 959 nm and 1972 nm, which are within 3 FSRs of the observed approximate DW peaks at 950 nm and 1965 nm. The calculated $\mathrm{D_{int}}$ for the polymer mask device has no zero crossings within a reasonable spectral span. We also observe a significantly larger curvature in $\mathrm{D_{int}}$ (i.e., larger anomalous dispersion) near the pump in this device, which might be connected to the higher threshold power for four-wave mixing. The differences in bandwidth and $\mathrm{D_{int}}$ between the two combs highlight the importance of controlling dimensions at the nm scale in realizing designed spectral breadth and shape. Furthermore, the angle of the sidewalls with respect to the substrate surfaces will also have a contribution to resonator dispersion, which will appear most significantly in the position of DWs. The positioning of the DWs in Fig. \ref{fig:MIcomb}(b) suggests that the sidewalls are indeed within 1\textdegree\,of normal; the $\mathrm{D_{int}}$ presented includes this sidewall angle, and even a change of 2\textdegree\,would lead to very different DW wavelengths (see Fig. S4 in the Supplemental Document for the exact DW dependence on resonator geometries calculated using FEM simulations).
For the polymer mask device comb in Fig. \ref{fig:MIcomb}(a), the lack of dispersive waves prevents us from probing the contributions of sidewall angle to the higher order terms in $\mathrm{D_{int}}$ at the same level. We attribute the absence of DWs in the comb spectrum generated in the polymer mask device to the difference in final resonator cross-sectional dimensions. We also observe soliton comb states (short-lived) in the comb power for devices fabricated using either process, shown in Fig. \ref{fig:MIcomb}(c) and (d) (from sample groups S2 and H1, respectively).

\section{Discussion and Conclusion}

We report here a series of experiments designed to establish process-structure-property (P-S-P) relationships connected to reducing optical loss and dispersion engineering in silicon nitride waveguides designed for nonlinear light generation. In these experiments, fabrication of nominally identical microresonators was carried out using near-identical fabrication processes and parameters, during which we changed only specific fabrication details that would logically have the greatest influence on scattering loss and resonator cross-sectional geometry. We investigated the influence of the fabrication processes by imaging the resonator physical structure (dimensions, sidewalls, sidewall angle) and measuring the optical properties (quality factor/optical loss, comb spectrum). 

In general, we find that using a metallic mask during etching leads to more faithful reproduction of rectangular cross-sections than a mask of polymer-based e-beam resist. We also find that the metal mask process results in increased sidewall roughness, potentially transferred from the metal mask itself. We supplement our understanding of the P-S-P relationships with two theoretical models. Firstly, we use the Payne-Lacey model to connect the measured optical loss with the sidewall roughness. Using this model, we estimate ranges for $\sigma$, the standard deviation of the sidewall corrugations, and $\mathrm{L_{C}}$, the correlation length of the deviations, associated with our level of propagation loss. 
Secondly, we use FEM simulations to model waveguide dispersions to verify the dimensions of the resonators based on their comb spectra. Our resonators are designed to have dual, harmonic dispersive waves, and the dispersive wave placement can be used to benchmark sidewall angle.

When developing a fabrication process for nonlinear optics in photonic circuits, it is important to define the most critical properties for the targeted application. In some cases, ultra-low scattering loss may be necessary for success, while in others, it may be secondary to phase-matching through nanometer accuracy in waveguide dimensions. In addition, in the setting of a multi-user academic cleanroom, individual users sometimes have limited access to and control over the condition of shared equipment, and time and effort must be directed towards the processes most directly influencing the critical device properties. 
It is our hope that this study suggests clear ways forward for researchers in this situation to benchmark and improve their fabrication results. While we have developed these processes in the well-studied material of silicon nitride, we believe that the results are transferrable to other nonlinear materials of interest. In particular, our focus on addressing problems posed by the etch-resistance of silicon nitride is relevant for developing subtractive fabrication for materials with even higher etch-resistivity, including tantalum pentoxide \cite{Jung2021} and lithium niobate \cite{Zhang2017}, both of which have promising applications in nonlinear photonic circuits \cite{Lamee2020,Snigirev2023}.

\section*{Funding}
We gratefully acknowledge financial support from AFOSR (FA9550-22-1-0174), NSF (2410813), NSF EPSCoR (2217786), and UNM Women in STEM awards.

\section*{Acknowledgment}
We thank scientists and staff of the Center for Integrated Nanotechnology (CINT) at Sandia National Labs, with particular thanks to John Nogan, Anthony James, and Michael De La Garza, and Denise Webb for their insightful and helpful discussions about fabrication process development. We are also grateful to Joseph Yelk for sharing his expertise in scientific illustration and for his contributions to Fig. \ref{fig:fig1}.

\section*{Disclosures}
The authors declare no conflicts of interest. 

\section*{Data availability}
The data for results presented in the paper and in the Supplemental Document is available upon reasonable request from the corresponding author. 

\section*{Supplemental document}
The Supplemental Document contains 6 sections and 4 figures and provides additional details about process fabrication, data presented in Table \ref{tab:loss}, the Payne-Lacey model, and the dependence of higher order terms in the dispersion (and dispersive wave placement) on resonator geometry.

\bibliography{references_v4}

\end{document}


\begin{abstract}   
\end{abstract}
\maketitle

\section{Silicon nitride deposition}

Stoichiometric silicon nitride (Si$_{3}$N$_{4}$) has a large amount of tensile stress that can cause cracking, particularly across uninterrupted films that are several inches across and with greater than 400 nm thickness. To protect our writeable area from cracks, we pattern trenches during the silicon oxidation process, prior to Si$_{3}$N$_{4}$ low pressure chemical vapor deposition (LPCVD). As a pre-oxidation measure, we deposit 100 nm of low-stress (silicon-rich) silicon nitride directly on the silicon wafer. This low stress nitride is then etched into a series of protective barriers in the outline of squares. After a very thorough RCA clean, the silicon is thermally oxidized, with the silicon dioxide forming a 3 $\upmu$m layer everywhere on the wafer aside from where the remaining nitride inhibits growth; these areas become trenches in the oxide, which transfers to trenches in the Si$_{3}$N$_{4}$ layer. These trenches act as crack barriers \cite{luke_overcoming_2013}, reducing the area over which the tensile stress accumulates and stopping any crack propagation from the wafer edges. The crack-barrier formation process was designed and developed with guidance and input from John Nogan, Sandia National Labs scientist and CINT Integration Lab manager.

Si$_{3}$N$_{4}$ deposition is a multi-step process, due to the high tensile stress of the material and the need to remove impurities. After oxidation, silicon nitride is deposited at 800 \textdegree C using low pressure chemical vapor deposition (LPCVD). This process uses a 1:3 ratio of dichlorosilane ($\mathrm{H_{2}SiCl_{2}}$) and ammonia ($\mathrm{NH_{3}}$) to deposit stoichiometric silicon nitride. We first deposit a 300 nm thick layer of silicon nitride, followed by an anneal at 1000 \textdegree C in a nitrogen atmosphere. The purpose of the anneal is to drive out hydrogen and eliminate residual N-H bonds created in the LPCVD process \cite{Henry1987-gn,luke_broadband_2015}. After annealing, we continue LPCVD until the desired silicon nitride thickness is achieved.

\section{Microresonator fabrication details}
This section outlines the fabrication process for both polymer and metal mask devices. The polymer mask process starts with the spin coating Si$_{3}$N$_{4}$ with a 1 $\mathrm{\upmu m}$ thick layer of ZEP520A, a positive-tone electron beam resist.
We expose the resist using JEOL JBX 6300-FS with an acceleration voltage of 100 kV, a beam current of 200 pA, beam step size of 4 nm, and the dose of exposure is 280 $\mathrm{\upmu C/cm^2}$. The exposed resist is developed using n-Amyl Acetate. This developed resist serves as the etch template for the silicon nitride waveguides.
The metal mask process begins with coating a 250 nm thick layer of ma-N 2403, a negative-tone electron beam resist. Resist samples are exposed using a beam current of 200 pA, a beam step size of 2 nm and the dose for exposure is 440 $\mathrm{\upmu C/cm^2}$. The exposed sample is then developed using MF 319. After development, we deposit a 50 nm layer of chromium using electron beam physical vapor deposition (EBPVD) and samples are soaked in \textit{N}-Methyl-2- pyrrolidone (NMP) to remove the resist and transfer the pattern in metal, which will serve as a etch resistant template \cite{colacion_low-loss_2025}. 

We etch resist and metal mask devices using inductively coupled plasma reactive ion etching (ICP-RIE) (Plasma-Therm SLR) using a mixture of carbon tetrafluoride ($\mathrm{CF_{4}}$), argon ($\mathrm{Ar}$), and oxygen ($\mathrm{O_{2}}$). The chemical component of this etch is same for both processes, but the physical component including the bias RF power is optimized to balance the trade-off between etch selectivity and directional etching. Due to the low etch selectivity of the polymer mask, a reduced bias RF power of 18 W is used, compared to 50 W for the metal mask. The larger bias RF power enabled by the metal mask etch-selectivity yields waveguides with nearly vertical sidewalls (as observed in Fig. 2 in the main text) and enhances etch-depth uniformity. On the other hand, the lower RF power of the polymer mask process leads to sloped sidewalls. After etching, the remaining  etch mask is removed. 
For edge coupling to our fabricated resonators, we use a combination of photolithography and etching to get smooth and low insertion loss facets \cite{ji_methods_2021}. The details of the facet release process are explained in Section \ref{sec:facets}. Finally, the Si$_{3}$N$_{4}$ resonators are annealed in a nitrogen atmosphere at 1180 \textdegree C for 3.5 hours to remove residual N-H bonds trapped in Si$_{3}$N$_{4}$ \cite{Henry1987-gn,luke_broadband_2015}.

\section{Etched facet process}
\label{sec:facets}

\begin{figure}[htpb]
    \centering
    \includegraphics[width=0.8\linewidth]{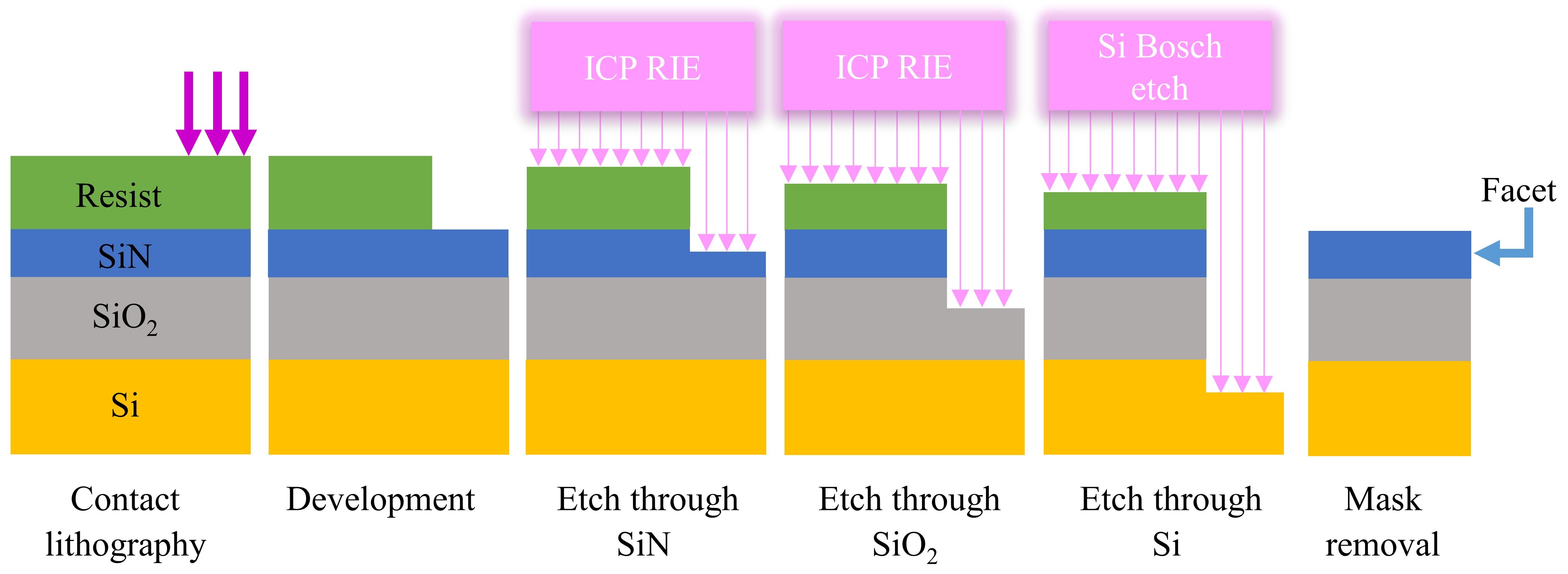}
    \caption{Schematic for the etched facet process using a combination of photolithography and etching.} 
    \label{fig:etched facet}
\end{figure}

The final step in device fabrication is typically the creation of facets for edge coupling of the light into the access waveguides (known as facet release). In general, waveguides can be cleaved or diced perpendicularly to release the facets for edge coupling, but this process can increase the insertion loss and we have found it inconsistent between chiplets. Therefore, in order to get more uniform insertion loss, we use photolithography followed by etching to release our facets \cite{ji_methods_2021}. Figure \ref{fig:etched facet} outlines the process for facet release. 

We start the process by spin coating our devices with AZ 10XT photoresist at 1300 rpm to create a 14 $\upmu$m thick layer of resist. This is followed by a 12 hour rehydration delay, which helps water diffuse into the resist from the atmosphere and leads to uniform development of the resist. We pattern the resist with trenches at the edge of the chip using a Karl Suss MA6 contact mask aligner with a dose of 600 $\mathrm{\upmu C/cm^2}$ over 30 seconds. (The edge of the sample is exposed twice for uniform development due to the ``edge beading'' effect.) The samples are developed using a combination of AZ 400K 1:4 and AZ 421K and rinsed with deionized (DI) water. After development, an $\mathrm{O_2}$ plasma descum process is performed for 1 minute to remove any residual undeveloped resist.

The developed samples are subject to a multi-step ICP-RIE. First, we etch the silicon nitride using a combination of $\mathrm{CF_4}$, $\mathrm{Ar}$, and $\mathrm{O_2}$. The etch rate for Si$_{3}$N$_{4}$ is 115 nm/min (silicon nitride is etched for approximately 5.5 min). The 3 $\upmu$m thick silicon dioxide ($\mathrm{SiO_2}$) bottom-cladding is etched using the same etch chemistry as silicon nitride, but with a higher bias RF power to increase the etch rate to 150 nm/min. We also incorporate thermal cycling to avoid burning of the resist. Finally, a Bosch etch is used to etch more than 200 $\upmu$m of silicon using a chemistry of $\mathrm{SF_6}$ and $\mathrm{C_4F_8}$. This process has a high etch selectivity of 35:1. The remaining resist is removed by soaking the samples in NMP overnight, followed by an $\mathrm{O_2}$ plasma clean. The finished samples are then rinsed using acetone and isopropanol. 

\section{Process limitation study using E-beam lithography}

Table 1 of the main text presents the average intrinsic quality factor and corresponding optical loss for both polymer and metal mask devices fabricated with varying beam current and shot pitch. In this section, we present the measured loaded quality factors and calculated $\mathrm{Q_{int}}$ for the individual resonances and devices that comprise these reported values.

\begin{figure}[ht!]
\centering
\includegraphics[width=1 \linewidth]{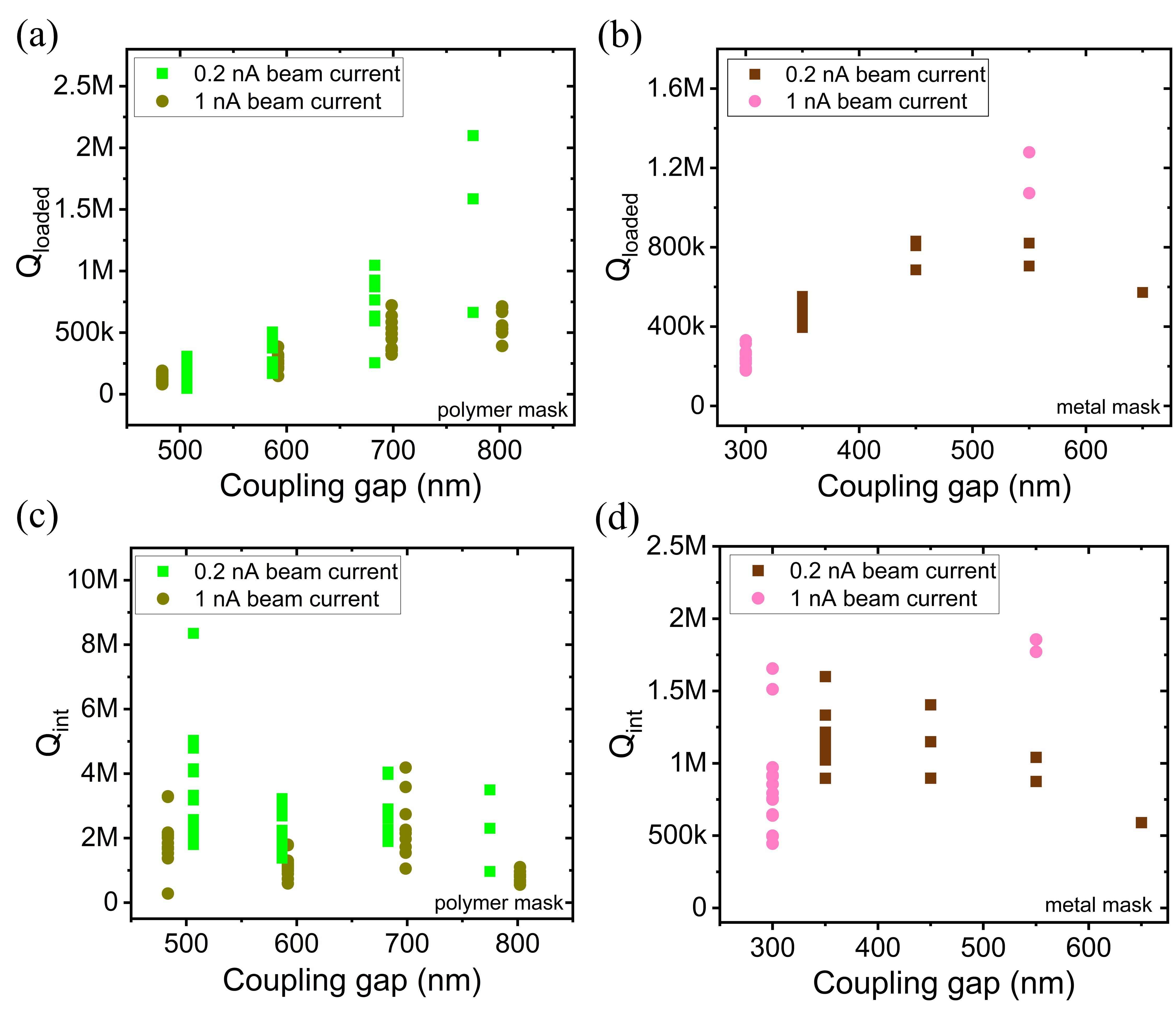}
\caption{{Effect of electron beam current on quality factor of fabricated resonators using polymer and metal mask process. Only singlet modes without interference from other mode families (mode crossings) are reported. (a) Loaded and (c) intrinsic quality factors vs. (measured) coupling gap using beam currents of 200 pA and 1 nA and the polymer mask process. We find that the devices patterned using a lower beam current (200 pA) have significantly higher $\mathrm{Q_{int}}$ than those patterned with higher beam current (1 nA). The shot pitch used for patterning these devices was kept at 8 nm for both beam currents. (b) Loaded and (d) intrinsic quality factors vs. (measured) coupling gap using beam currents of 200 pA and 1 nA and the metal mask process. We do not observe a significant dependence on beam current for these devices. The shot pitch used for patterning these devices was kept at 6 nm for both beam currents. (Panels b and d are adapted from reference \citenum{colacion_low-loss_2025}.)}}
\label{fig:fig5}
\end{figure}

Figure \ref{fig:fig5} presents loaded and intrinsic quality factors in devices with multiple coupling gaps for variations in beam current. Keeping the shot pitch fixed, we set the current of the electron beam during lithography to be either 200 pA or 1 nA, which corresponds to expected beam diameters of 5.8 nm and 7.5 nm at the resist surface, respectively. Figures \ref{fig:fig5}(a) and \ref{fig:fig5}(c) show results for polymer mask devices. We identify critical coupling between 700 nm and 800 nm, and we observe that as we move towards undercoupled regime (larger gaps) the difference between $\mathrm{Q_{loaded}}$ between 200 pA and 1 nA becomes more significant, which is consistent with the enhancement of $\mathrm{Q_{int}}$ for the lower e-beam current. We find an average $\mathrm{Q_{int}}$ for 200 pA beam current of 2.8(2)M and an average $\mathrm{Q_{int}}$ of 1.6(1)M for 1 nA. This corresponds with an increase in propagation loss of 0.13 dB/cm to 0.23 dB/cm. 
Figures \ref{fig:fig5}(b) and \ref{fig:fig5}(d) present results for the metal mask devices. We calculate an average $\mathrm{Q_{int}}$ of 1.10(7)M for 200 pA and  1.0(1)M for 1nA beam current, which correspond to propagation loss of 0.34 dB/cm and 0.38 dB/cm, respectively. We report no significant dependence of optical loss on beam current for these devices. We find that critical coupling for this set of metal mask devices lies between gaps of 300 and 400 nm, which is consistent with both the lower overall $\mathrm{Q_{int}}$ and our observation of well-etched gap regions for the metal mask process.

\begin{figure}[ht!]
\centering
\includegraphics[width=1\linewidth]{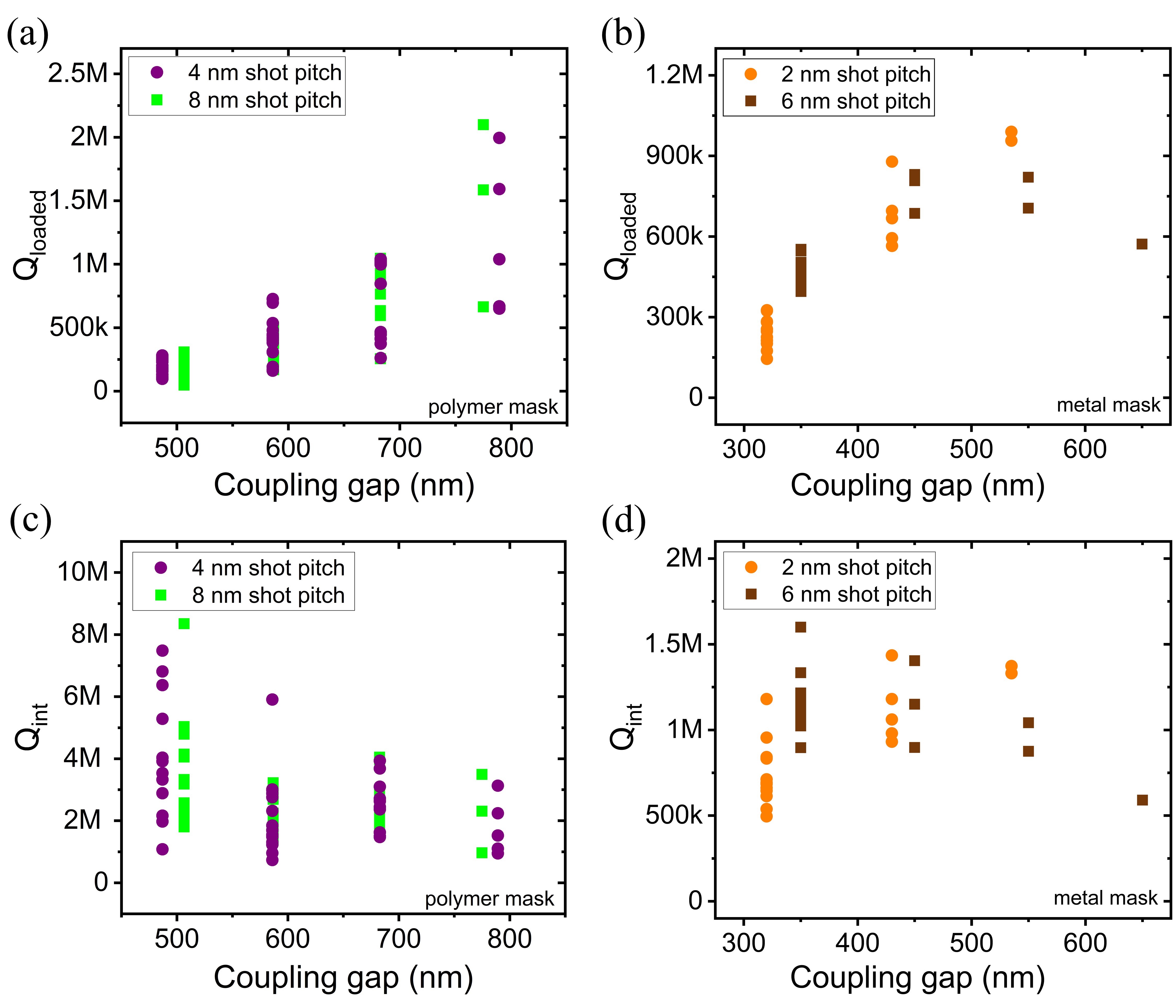}
\caption{Effect of shot pitch on the quality factors of resonators fabricated using the polymer and metal mask processes. The electron beam current was 200 pA for all data in these panels. (a) Loaded and (c) intrinsic quality factors vs. (measured) coupling gap for resonators fabricated uing the polymer mask process. The EBL shot pitch was varied from 4 nm to 8 nm. (b) Loaded and (d) intrinsic quality factors vs. (measured) for resonators fabricated using the metal mask process. The EBL shot pitch was varied from 2 nm to 6 nm. (Panels b and d are adapted from reference \citenum{colacion_low-loss_2025}.)}
\label{fig:figure4}
\end{figure} 

Figure \ref{fig:figure4}(a) and \ref{fig:figure4}(c) compare the loaded and intrinsic quality factors for shot pitches of 4 nm and 8 nm and a beam current 200 pA for devices fabricated using an etch-mask of polymer-based resist. We calculate an average $\mathrm{Q_{int}}$ for polymer mask devices of 2.7(2)M and 2.8(2)M for shot pitches of 4 nm and 8 nm, respectively. This corresponds to propagation loss of 0.14 dB/cm and 0.13 dB/cm; we do not find a statisically significant trend in this data.
Similarly, Fig. \ref{fig:figure4}(b) and \ref{fig:figure4}(d) compare quality factors in metal mask process devices with shot pitch variation between 2 nm and 6 nm (e-beam current us 200 pA). The average calculated $\mathrm{Q_{int}}$ values are 0.90(7)M and 1.10(7)M for 2 nm and 6 nm shot pitch, with a propagation loss of 0.42 dB/cm and 0.34 dB/cm respectively. Like in the polymer mask process, we do not find a connection between optical loss and EBL shot pitch for the metal-mask processed resonators. 

\section{Payne-Lacey Model for Scattering Loss}

The analytical expression first conceived by F. P. Payne and J. P. R. Lacey to describe scattering loss in planar optical waveguides is given by \cite{Payne_1994}
\begin{equation}
    \alpha_{\mathrm{scattering}} = 4.34\,\frac{\sigma^2}{\sqrt{2} \, k_0 \, d^4 \, n_{\mathrm{core}}}\,g(V)\,f(x,\gamma),
    \label{Eq:PayneLacey}
\end{equation}
where we've multiplied the scattering loss coefficient $\alpha$ by a factor of 4.34 to give the loss in units of $\mathrm{dB/m}$. The scattering loss in Equation \ref{Eq:PayneLacey} is written in terms of the standard deviation of the roughness ($\sigma$), the free space wavenumber ($k_0$), the half-width of the waveguide ($d$), and the refractive index of the core ($n_{\mathrm{core}}$) as well as functions $g(V)$ and $f(x,\gamma)$. 
The function $g(V)$ has the form
\begin{equation}
    g(V) = \frac{U^2\,V^2}{1+W}
    \label{Eq:g}
\end{equation}
and is purely a function of the waveguide materials and geometry with:
\begin{equation}
    U = k_0d\,\sqrt{n_{\mathrm{core}}^2 - n_{\mathrm{eff}}^2}\,
    \label{Eq:U}
\end{equation}
\begin{equation}
    V = k_0d\,\sqrt{n_{\mathrm{core}}^2 - n_{\mathrm{clad}}^2}\,
    \label{Eq:V} 
\end{equation}
\begin{equation}
    W = k_0d\,\sqrt{n_{\mathrm{eff}}^2 - n_{\mathrm{clad}}^2}\,
    \label{Eq:W}
\end{equation}
where $n_{\mathrm{clad}}$ is the waveguide cladding material refractive index and $n_{\mathrm{eff}}$ is the effective index of refraction for the waveguide geometry. The effective index of refraction is determined using a finite-element method solver (COMSOL Multiphysics\textregistered) for a Si$_{3}$N$_{4}$ ring resonator with \SI{23.3}{\micro\meter} radius, \SI{1820}{\nano\meter} ring width, and \SI{620}{\nano\meter} ring height. The calculated effective index at our analysis wavelength of \SI{1550}{\nano\meter} is 1.7836, whereas the bulk Si$_{3}$N$_{4}$ index of refraction is 1.9977. The cladding index is 1, as the waveguide is air-clad.
The function $f(x,\gamma)$ has the form
\begin{equation}
    f(x,\gamma) = \frac{x\sqrt{1-x^2+\sqrt{(1+x^2)^2 + 2 x^2 \gamma^2}}}{\sqrt{(1+x^2)^2 + 2 x^2 \gamma^2}}
    \label{Eq:f}
\end{equation}
where 
\begin{equation}
    x = W \frac{L_{\mathrm{C}}}{d}
    \label{Eq:x}
\end{equation}
\begin{equation}
    \gamma = \frac{n_{\mathrm{clad}} V}{n_{\mathrm{core}} W \sqrt{\Delta}}
    \label{Eq:gamma}
\end{equation}
\begin{equation}
    \Delta = \frac{n_{\mathrm{core}}^2 - n_{\mathrm{clad}}^2}{2 n_{\mathrm{core}}^2}
    \label{Eq:delta}
\end{equation}
and $L_{\mathrm{C}}$ is the correlation length of the roughness. Given the waveguide dimensions and composition, $L_{\mathrm{C}}$ and $\sigma$ together estimate the scattering loss, $\alpha_{\mathrm{scattering}}$.

\section{Impact of dimensions on resonator dispersion}

Waveguide dimensions play a crucial role in determining a resonator's dispersion \cite{okawachi_octave-spanning_2011,tan_group_2010}. The geometry of the resonator is tailored to overcome the material normal dispersion and create a window of anomalous dispersion and to extend the comb bandwidth via dispersive waves. The fabrication process can complicate achieving the target waveguide dimensions, which in turn affects the ability of the fabricated devices to generate broadband octave-spanning frequency combs with well-positioned dispersive waves. Expanding on Fig. 5 and the related discussion in the main text, we examine the impact of changes in width, height, and sidewall angle on the position of the dispersive waves. 

Finite element method COMSOL simulations are used to calculate the dispersion of a Si$_{3}$N$_{4}$ ring resonator. First, we calculate the effective index of an air-clad Si$_{3}$N$_{4}$ resonator. We simulate a ring resonator with a measured cross section of 1730 nm by 582 nm with straight, vertical sidewalls. The effective index is used to calculate the comb mode index ($m$) as a function of frequency and the integrated dispersion, which is the deviation of resonant frequencies from an equidistant grid. The integrated dispersion, $\mathrm{D_{int}}$, is defined as
\begin{equation}
    \mathrm{D_{int}} = \omega_{\mu}-\omega_\mathrm{{0}}-\mu*\mathrm{FSR},
    \label{Eq:Dint}
\end{equation}
where $\omega_{\mu}$ is the resonant frequency of the mode $\mu=m-m_{0}$, $\omega_\mathrm{{0}}$ is the pumped resonance frequency with index $m_{0}$, and FSR is the free spectral range. The presence of higher-order dispersion terms gives rise to the formation of dispersive waves. The spectral location of dispersive waves (DWs) occurs when  $\mathrm{D_{int}}(\mu)=0$ 
far from the pump \cite{Brasch_2016}. 

\begin{figure}[ht]
    \centering
    \includegraphics[width=1\linewidth]{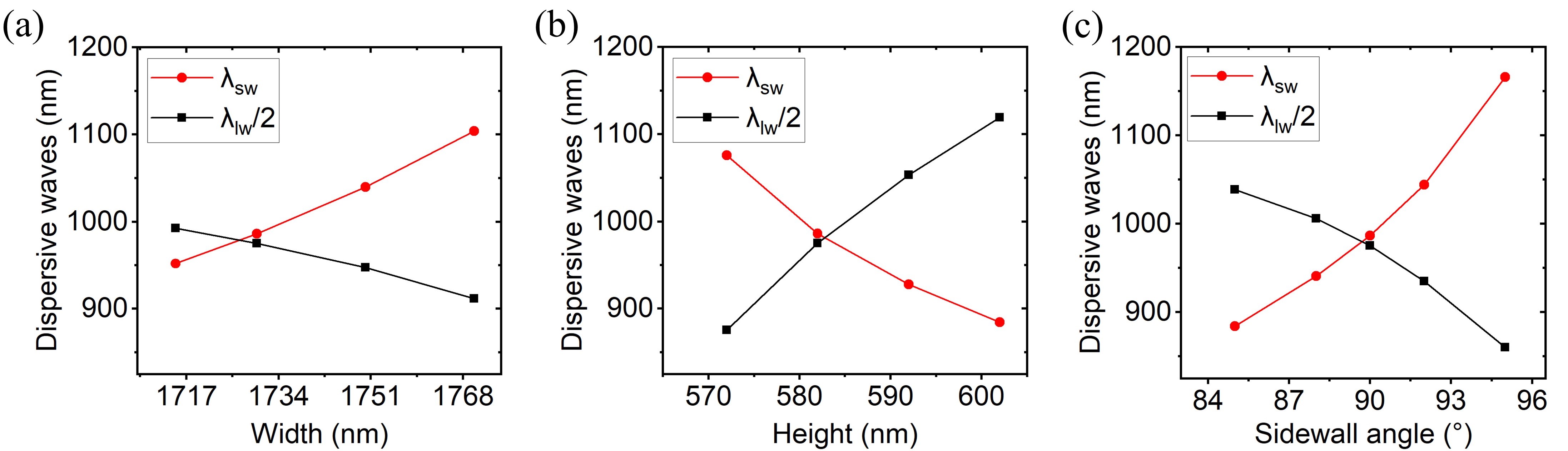}
    \caption{Impact of geometry on the spectral location of dispersive waves. (a) Spectral location of short wavelength ($\lambda_\mathrm{_{sw}}$)  and long wavelength dispersive waves ($\lambda_\mathrm{_{lw}}$) as a function of resonator waveguide width. (b) Spectral location of $\lambda_\mathrm{_{sw}}$ and $\lambda_\mathrm{_{lw}}$ as a function of waveguide height. (c) Spectral location of $\lambda_\mathrm{_{sw}}$ and $\lambda_\mathrm{_{lw}}$ as a function of sidewall angle (90\textdegree\, is vertical).  
    } 
    \label{fig:dim_disp}
\end{figure}

Figure \ref{fig:dim_disp} shows how varying the geometry of the resonator (width, height, and sidewall angle) affects the spectral location of DWs. To identify the trend in DWs as a function of geometry, we fit the data to a line and estimate the approximate change in DW wavelength. At a constant Si$_{3}$N$_{4}$ thickness of 582 nm and vertical sidewalls, an increase in the width of the resonator causes both short and long wavelength DWs to move closer to the pump wavelength (Fig. \ref{fig:dim_disp}(a)). Similarly, at a constant width of 1730 nm and vertical sidewalls, an increase in Si$_{3}$N$_{4}$ thickness causes both DWs to move away from the pump. 

We find that, for the spectral position of both the low energy dispersive wave peak ($\lambda_\mathrm{_{lw}}$) and the high energy dispersive wave peak ($\lambda_\mathrm{_{sw}}$), the wavelengths are twice as susceptible to changes in height as to changes in width (i.e., the fitted slope is twice as large). 
Figure \ref{fig:dim_disp}(c) shows the spectral location of DWs as a function of sidewall angle (1730 nm by 582 nm cross section). We observe that the position of the dispersive waves is sensitive to sidewall angle; even a change of 1\textdegree\, will steer $\lambda_\mathrm{_{sw}}$ four times farther than a 1 nm change in waveguide height on average.  
We target the condition $\lambda_\mathrm{_{sw}}$ = $\lambda_\mathrm{_{lw}}/2$ (harmonic DWs), which facilitates comb stabilization via f-2f self-referencing \cite{drake_terahertz-rate_2019}.

\bibliography{supplemental_references}